\documentclass{article}

\usepackage[preprint]{spconf}

\usepackage{hyperref}

\usepackage[T1]{fontenc} % add special characters (e.g., umlaute)
\usepackage[utf8]{inputenc} % set utf-8 as default input encoding
\usepackage{spconf,amsmath,cite,url}
\usepackage{microtype}
\usepackage{xfrac}
\usepackage{graphicx}
\usepackage{color}
\usepackage[table,xcdraw]{xcolor}
\usepackage{hhline}
\usepackage{multirow}
\usepackage{eso-pic}
\usepackage{tikz}

\newcommand {\minisec}[1]{\noindent\textbf{#1.}}

\newcommand\copyrighttext{%
  \footnotesize Copyright 2022 IEEE. Published in ICASSP 2022 - 2022 IEEE International Conference on Acoustics, Speech and Signal Processing (ICASSP), scheduled for 7-13 May 2022 Virtual; 22-27 May 2022 In-Person in Singapore. Personal use of this material is permitted. However, permission to reprint/republish this material for advertising or promotional purposes or for creating new collective works for resale or redistribution to servers or lists, or to reuse any copyrighted component of this work in other works, must be obtained from the IEEE. Contact: Manager, Copyrights and Permissions / IEEE Service Center / 445 Hoes Lane / P.O. Box 1331 / Piscataway, NJ 08855-1331, USA. Telephone: + Intl. 908-562-3966.}
\newcommand\copyrightnote{%
\begin{tikzpicture}[remember picture,overlay]
\node[anchor=south,yshift=10pt] at (current page.south) {\fbox{\parbox{\dimexpr\textwidth-\fboxsep-\fboxrule\relax}{\copyrighttext}}};
\end{tikzpicture}%
}

\ninept

\title{A Lightweight Instrument-Agnostic Model for Polyphonic Note Transcription and Multipitch Estimation}

\name{Rachel M. Bittner$^{\flat}$, Juan Jos\'e Bosch$^{\flat}$, David Rubinstein$^{\flat}$, Gabriel Meseguer-Brocal$^{\sharp}$, Sebastian Ewert$^{\flat}$}
\address{$^{\flat}$Spotify, $^{\sharp}$IRCAM}% Inc.}

\begin{document}

\maketitle

\begin{abstract}

Automatic Music Transcription (AMT) has been recognized as a key enabling technology with a wide range of applications. Given the task's complexity, best results have typically been reported for systems focusing on specific settings, e.g. instrument-specific systems tend to yield improved results over instrument-agnostic methods.
Similarly, higher accuracy can be obtained when only estimating frame-wise $f_0$ values and neglecting the harder note event detection. Despite their high accuracy, such specialized systems often cannot be deployed in the real-world. Storage and network constraints prohibit the use of multiple specialized models, while memory and run-time constraints limit their complexity.
In this paper, we propose a lightweight neural network for musical instrument transcription, which supports polyphonic outputs and generalizes to a wide variety of instruments (including vocals).
Our model is trained to jointly predict frame-wise onsets, multipitch and note activations, and we experimentally show that this multi-output structure improves the resulting frame-level note accuracy. 
Despite its simplicity, benchmark results show our system's note estimation to be substantially better than a comparable baseline, and its frame-level accuracy to be only marginally below those of specialized state-of-the-art AMT systems.
With this work we hope to encourage the community to further investigate low-resource, instrument-agnostic AMT systems.
\end{abstract}

\copyrightnote


\section{Introduction}\label{sec:introduction}

The automatic transcription of music has been studied for more than four decades \cite{benetos_automatic_2019}.
%\cite{moorer1977transcription, benetos_automatic_2019}. 
During this time, systems have considerably improved, in particular since the rise of deep learning. 
Yet, the task remains unsolved, partially due to various intrinsic challenges \cite{benetos_automatic_2019} but also due to a lack of an objective ground truth on which humans consistently agree \cite{su2015escaping}. 
Because of the intrinsic difficulty of the task, AMT systems are usually designed with a limited scope, and focus on a sub-task.
There are a number of common sub-tasks in AMT which branch along three dimensions: (1) the degree of output polyphony (monophonic, polyphonic) (2) the types of output to be estimated (notes, $f_0$), and (3) the type of input audio (pop songs, solo piano, solo guitar, jazz ensembles, etc.).
% To improve results by simplifying the problem, the AMT task is often further subdivided into various sub-tasks or  specializations. 
For example, specializing for a specific instrument class allows models to exploit instrument-specific characteristics to increase the transcription accuracy, e.g. piano \cite{sigtia_end--end_2016, hawthorne_enabling_2019, hawthorne2021sequence}, guitar~\cite{su_tent_2019,wiggins_guitar_2019} or singing voice~\cite{mcleod2017automatic, hsu_vocano}.
Similarly, models built to estimate a particular output type, or which are restricted to monophonic settings\cite{kim_crepe_2018} can further increase accuracy in these scenarios.
In many real-world applications, deploying a number of specialized systems becomes intractable, for example because of storage, network and maintenance constraints.
Further, for many instruments it is challenging to create a dataset large enough to train modern methods.
Applications can also add additional restrictions w.r.t. the size of the model, its (peak) memory consumption and run-time. 
Therefore, there is often a gap between the latest published state of the art and models that can practically be deployed in a range of settings.

In this work, we consider a broad scenario: an  instrument-agnostic\footnote{By ``instrument agnostic'' we mean ``not specific to an instrument class''.} polyphonic AMT model which estimates both notes and multipitch outputs.
The proposed model is a lightweight neural network which runs efficiently on low-end devices, thanks to its low memory and processing time requirements. 
Unless otherwise noted, we deal with polyphonic recordings of a single instrument class (e.g. solo piano, an ensemble of violins, solo vocals, a choir, etc.), but do not restrict which classes we consider.
It is jointly trained to predict frame-level onset, multipitch and note posteriorgrams.
During inference, we post-process the frame-level posteriorgrams to obtain note events and multipitch information.
We study the ability of the proposed model to transcribe a variety of instruments and vocals 
% in single-instrument polyphonic musical audio 
without retraining, and compare with a recent baseline model for instrument-agnostic polyphonic note estimation.
Further, we evaluate the contribution of components of the proposed model with an ablation study.
All code and trained models discussed in this paper are made publicly available\footnote{\url{https://github.com/spotify/basic-pitch}}.
Additionally, we only use public datasets for training and evaluation in order to foster reproducibility. 

\section{Background and Related Work}\label{sec:relatedwork}

There is a huge body of work on AMT. Due to space constraints we refer to \cite{benetos_automatic_2019, klapuri2007signal} for a more comprehensive overview.
As previously mentioned, AMT systems have three dimensions: (1) the degree of output polyphony considered, (2) the type of output estimated and (3) the type of input audio.
In this work we consider the polyphonic setting, where more than one note/pitch may be present in the output at a time; note that monophonic AMT is a strict subset of polyphonic AMT, and thus we also support monophonic sources.
AMT outputs are typically either \textbf{frame-level multipitch estimation (MPE)} or \textbf{note-level estimation}, which transcribe polyphonic music at different levels of granularity\cite{benetos_automatic_2019}.
Both are useful depending on the application: MPE provides lower-level expressive performance information (such as vibrato, glissando), whereas note-level estimation gives information closer to the musical score.
MPE methods predict the fundamental frequencies ($f_0$s) which are active at a given time-frame (note that even if not strictly equivalent, we use pitch and $f_0$ interchangeably following the literature in the field \cite{benetos_automatic_2019}).
They commonly first estimate a pitch posteriorgram~\cite{duan2010multiple,bittner_deep_2017}, where each time-frequency bin is assigned an estimate of the likelihood of that fundamental frequency being active at a given time. 
Such matrices typically contain multiple bins per semitone, which allows estimation of small (``continuous'') variations of pitch. 
Various methods aim at estimating and subsequently grouping MPE outputs from polyphonic recordings into note events \cite{ryynanen2005polyphonic,duan2013multi,benetos2017polyphonic,ycart2018polyphonic,nishikimi_bayesian_2020}, or attempt to group pitches into contours~\cite{klapuri2007signal,duan2010multiple,bittner_pitch_2017}. 
Note estimation (or note tracking) methods aim at estimating notes events (defined as: pitch, onset time, offset time). %, which could be used to reproduce the recording on another instrument.
Notes cannot be trivially estimated from the output of an MPE system, because MPE information does not encode onsets/offsets, and preserves fluctuations in pitch which should not always be quantized to the nearest semitone.
In particular, note estimation is difficult for singing voice, which may have a high degree of fluctuation around a center pitch~\cite{hsu_vocano} compared to instruments such as the piano.
%This task is far from trivial, as it involves grouping frame-level activations in both time and frequency.
Multiple methods have been proposed for estimating notes from pitch posteriorgrams e.g. using median filtering \cite{klapuri2007signal}, Hidden Markov Models \cite{benetos2017polyphonic} or neural networks \cite{EwertS2017_PianoTransADMM+LSTM_WASPAA,nishikimi_audio--score_2021}. 
While most approaches consider each semitone independently, some approaches attempt to model the interactions between notes, using spectral likelihood models \cite{nishikimi_bayesian_2020,benetos_automatic_2019}, or music language models \cite{sigtia_end--end_2016,ycart2018polyphonic}. Transformers have recently been applied to AMT, directly predicting MIDI-like note events from spectrograms in piano music \cite{hawthorne2021sequence}. 
A few AMT models perform both note and pitch estimation ~\cite{mauch_computer-aided_nodate, ryynanen2005polyphonic, nishikimi_bayesian_2020}, and most work with monophonic data. %Most of them perform pitch estimation as a pre-processing step for note estimation. 
Regarding input audio characteristics, traditional AMT methods based on signal-processing have been more generalizable to multiple instruments than more recent approaches, as well as being simpler and faster\cite{benetos_automatic_2019,duan2010multiple}. 
However, the best performing systems often come at the expense of higher computational requirements and a focus on instrument-specific systems\cite{hawthorne_enabling_2019}.

\section{Model}

Our goal is to create an AMT model that generalizes across a set of polyphonic (or monophonic) instruments without retraining, while being lightweight enough to run in low-resource settings.
We consider both the speed and the peak memory usage when running inference, and purposely limit ourselves to a shallow architecture to keep the memory needs low.
Note that the number of parameters of a model does not necessarily correlate with its memory usage; e.g. a convolution layer requires few parameters, but can still have high memory usage due to the feature map sizes.

\minisec{Harmonic Stacking}
Given the input audio, the model first computes a Constant-Q Transform (CQT) with 3 bins per semitone and a hop size of $\approx$ 11 ms.
Rather than using, e.g. a mel spectrogram and ultimately learning the projection onto the output log-spaced frequency scale using a Dense or LSTM layer (which requires the model to have a full-frequency receptive field)\cite{hawthorne_enabling_2019}, we 
start with a representation with the desired frequency scale.
The Harmonic CQT (HCQT)~\cite{bittner_deep_2017} is a transformation of the CQT which aligns harmonically-related frequencies along a third dimension, allowing small convolutional kernels to capture harmonically-related information. 
As an efficient approximation of the HCQT, following~\cite{balhar_melody_nodate},
we copy the CQT and shift it vertically by the number of frequency bins corresponding to each harmonic. In this work we use 7 harmonics and 1 sub-harmonic.

\begin{figure}
    \centering
    \includegraphics[width=0.88\columnwidth]{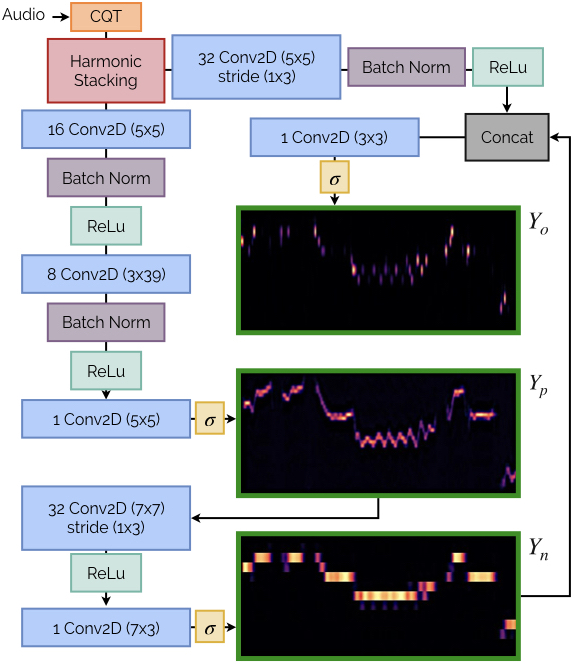}
    \caption{The NMP architecture. The matrix posteriorgram outputs $Y_o$, $Y_p$, and $Y_n$ are outlined in green. $\sigma$ indicates a sigmoid activation.}
    \label{fig:architecture}
\end{figure}

\minisec{Architecture} 
The architecture illustrated in Fig. ~\ref{fig:architecture} is a fully convolutional model taking audio as input and produces three posteriorgram outputs, with a total of only 16,782 parameters.
The model's three output posteriorgrams are time-frequency matrices encoding if (1) the onset of a note is taking place ($Y_o$) (2) a note is active ($Y_n$) and (3) a pitch is active ($Y_p$).
All outputs have the same number of time frames as the input CQT, and in frequency, both $Y_o$ and $Y_n$ have a resolution of 1 bin per semitone while $Y_p$ has a resolution of 3 bins per semitone.
Besides having different frequency resolutions, $Y_n$ and $Y_p$ are trained to capture different concepts: $Y_n$ captures frame-level note event information ``musically quantized'' in time and frequency, while $Y_p$ encodes frame level multipitch information, capturing details such as vibrato.
During training, the target for each of these outputs are binary matrices generated from note and pitch annotations.

The architecture is structured in order to exploit the differing properties of the three outputs.
We assume that $Y_p$ is the output which is ``closest'' to the input audio.
The architecture estimating $Y_p$ is similar to that of \cite{bittner_deep_2017}, but with fewer convolutional layers to reduce memory usage.
Notably, we employ the same octave plus one semitone-sized kernel in frequency, which we found to be helpful for avoiding octave mistakes.
This stack of convolutions performs a form of ``denoising'', in order to emphasize the multipitch posterior outputs and de-emphasize transients, harmonics and other un-pitched content.
An added benefit to using a limited receptive field in frequency is that it removes the need for pitch shifting data augmentations.
$Y_p$ followed by two small convolutional layers are used to estimate $Y_n$.
These convolutions can be seen as ``musical quantization'' layers, learning how to perform the non trivial grouping of multipitch posteriorgrams into note event posteriorgrams.
Finally, as in~\cite{hawthorne_onsets_2018}, $Y_o$ is estimated using as input both $Y_n$ and convolutional features computed from the audio, which are necessary to identify transients.

\minisec{Training}
Binary cross entropy is used as the loss function for each output, and the total loss is the sum of the three losses.
However, for $Y_o$, there is a heavy class imbalance that drives models to output $Y_o=0$ everywhere.
As a countermeasure, we use a class-balanced cross entropy loss, where the weight for the negative class is 0.05 and the positive is 0.95 (set empirically by observing the properties of the resulting $Y_o$) which helps the model capture the onsets while remaining sparse.
During training, the model input is 2 seconds of audio at a sample rate of 22050 Hz. We train the model with a batch size of 16 and use the Adam optimizer with a learning rate of 0.001.
During training, random label-preserving augmentations are applied to the audio, including adding noise, equalization filters, and reverb.

\minisec{Posteriorgram post-processing}
Similar to many note or contour creation post-processing methods, we create note events, defined by a start time $t^0$, end time $t^1$ and a pitch $f$ by running a post-processing step using $Y_o$ and $Y_n$ as inputs \cite{benetos_automatic_2019}, following a process similar to that described in Onsets and Frames~\cite{hawthorne_enabling_2019}.
A set of onset candidates $\{(t^0_i, f_i) \}$ are populated by peak picking $Y_o$ across time, and discarding peaks with likelihood $< 0.5$.
Note events are created for each $i$ in descending order of $t^0_i$, by tracking forward in time through $Y_n$ until the likelihood falls below a threshold $\tau_n$ for longer than an allowed tolerance (11 frames), then ending the note.
When notes are created, the likelihood of all corresponding frames of $Y_n$ are updated to 0.
After all onsets have been used, additional note events are created by iterating through bins of $Y_n$ that have likelihood $>\tau_n$ in descending order, following the same note creation procedure but instead tracing both forward and backward in time.
Finally, note events which are shorter than $\approx 120$ ms are removed.
Multi-pitch estimates are created by simply peak picking $Y_p$ across frequency and retaining all peaks greater than $\tau_n$.

\section{Experiments}

In this section we examine the performance of the proposed method, ``Notes and Multipitch'' (NMP), focusing on the note estimation task, but also briefly commenting on the MPE task. 
AMT methods have commonly been evaluated using a set of metrics proposed for MIREX\footnote{\url{http://www.music-ir.org/mirex/}} evaluation tasks.
In this work we report the note-level F-measure (\texttt{F}), where notes are considered correct if the pitch is within a quarter tone, the onset is within 50 ms, and the offset is within 20\% of the note's duration, the note-level F-measure-no-offset (\texttt{Fno}) with the same criterion as F-measure, but ignoring offsets, and the frame-level note accuracy (\texttt{Acc}), which is computed for frames with a hop size of 10 ms.
We use \texttt{Fno} as the main measure of overall note estimation accuracy since the definition of offsets is less objective than onsets (e.g. due to reverberation, sustain pedal, annotation procedure)~\cite{liang2015musical}.
We compute these metrics using \texttt{mir\_eval}~\cite{raffel2014mir_eval}.
For NMP and each of the ablation studies, we fine-tune the note creation parameter $\tau_n$ on the validation dataset such that it maximizes \texttt{Fno}.

In order to assess how well NMP and the baseline perform across different instrument classes, we use a wide variety of training and test data spanning multiple instrument types, summarized in Table~\ref{tab:datasets} (see the cited papers for more specific details), using the ~\texttt{mirdata} \cite{mirdata_zenodo} library.
A random 5\% of tracks from the training set are used for validation. We note a few additional details for some the datasets: We use the de-duplicated ``redux'' version of Slakh, and test on an instrument-balanced subset of 120 of the non-percussive test-set stems with the least silence;
the note annotations in MedleyDB and iKala are automatically generated using pyin-notes~\cite{mauch_computer-aided_nodate};
the audio files for MedleyDB are taken from the pitch tracking subset\footnote{\url{https://zenodo.org/record/2620624}}, and for iKala, we use the isolated vocals;
for Phenicx, we use the 42 instrumental section-grouped stems (e.g. violins, bassoons) and annotations.

\begin{table}
\small
\centering
\resizebox{\columnwidth}{!}{\begin{tabular}{|r|c|c|c|c|c|} 
\hline
\textbf{Dataset} & \textit{\textbf{Polyphony}} & \textit{\textbf{Instrument}}& \textit{\textbf{Labels}}  & \textit{\textbf{Train}\textbf{}} & \textit{\textbf{Test}}  \\ 
\hline
\textit{Molina} \cite{molina2014evaluation} & Mono & Vocals & N & -  & 38 \\
\hline
\begin{tabular}[r]{@{}c@{}}\textit{GuitarSet} \cite{xi2018guitarset}  \end{tabular} & Mono / Poly & \begin{tabular}[c]{@{}c@{}}Ac. Guitar\end{tabular} & N + P & 648 & 72 \\
\hline
\textit{MAESTRO} \cite{hawthorne_enabling_2019} & Poly & Piano  & N & 1154  & 128\\
\hline
\textit{Slakh} \cite{manilow2019cutting}  & Poly & Synthesizers  & N & 1590  & 120\\
\hline
\textit{Phenicx} \cite{miron2016score} & Poly & Orchestral& N & -  & 42 \\
\hline
\textit{iKala} \cite{chan2015vocal}  & Mono & Vocals & N + P & 252& -  \\
\hline
\textit{MedleyDB}  \cite{bittner2014medleydb} & Mono & Multiple  & N + P & 103& -  \\
\hline
\end{tabular}}
\caption{\small Summary of the datasets used. The \textit{Train} and \textit{Test} columns indicate the number of tracks. The \textit{Labels} column indicates which kind of annotations are available: (N) Notes, (P) Multi-pitch.}
\label{tab:datasets}
\end{table}

\subsection{Note Transcription Baseline Comparison}

We compare our model with a recent, strong baseline model, \textbf{MI-AMT}\cite{wu_multi-instrument_2020} which is a polyphonic, instrument-agnostic note estimation method. 
It uses a U-Net architecture with an attention mechanism and outputs a note-activation posteriorgram
with a total of over 20M parameters, trained on MAESTRO and MusicNet. The note posteriorgram is post-processed in order to create note events.

Results for our proposed method and MI-AMT are presented in Table \ref{tab:results}. 
We first remark that NMP considerably outperforms the baseline MI-AMT on all test datasets and metrics, with the exception of the comparable \texttt{Acc} on MAESTRO (piano) and Slakh (synthesizers). NMP performs strongly for datasets with polyphonic instruments (MAESTRO, Slakh, Phenicx, \sfrac{1}{2} of GuitarSet) as well as monophonic (Molina and \sfrac{1}{2} of GuitarSet) despite not imposing a monophonic constraint on the output note estimates. Additionally, we see consistent performance across datasets with varying instrument types, validating that NMP performs well without needing to be instrument-specific.

\begin{table}[ht]
    \centering
    \includegraphics[width=\columnwidth]{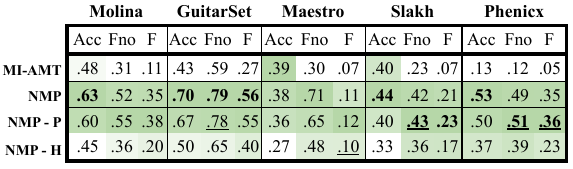}
    % \vspace*{-7mm}
    \caption{\small Average note event metrics on all test datasets for the baseline algorithm, proposed method, and ablation experiments. The best score for each column is in bold. The shade of green indicates how far a score is from the best score, with the worst scores in white. 
    All non-underlined results are statistically significantly different with $p<0.05$ compared with NMP (per metric/dataset) via a paired t-test.}
    \label{tab:results}
\end{table}

\subsection{Ablation Experiments}

\minisec{Harmonic Stacking} To examine the use of Harmonic Stacking as an input representation, we trained a model which omits the harmonic stacking layer but is equivalent otherwise, denoted as \texttt{NMP - H} in Table~\ref{tab:results}.
Unsurprisingly, given the small receptive field, the omission of harmonic stacking substantially reduces performance across all metrics and datasets, in accordance with the results of similar experiments performed in~\cite{bittner_deep_2017,balhar_melody_nodate}.
This indicates that Harmonic Stacking effectively allows the model to use smaller convolutional kernels while still capturing relevant information. One limitation of this comparison is that the number of channels is reduced when omitting harmonic stacking, which in turn reduces the model's capacity.

\minisec{The Effect of $Y_p$} We measure the impact that the supervised bottleneck layer $Y_p$ on $Y_n$ has on note estimation by training an equivalent model, where $Y_p$ is not supervised, and where $Y_n$ is the output of the stack of convolutions preceding it, with the \textit{Batch Norm} $\rightarrow$ \textit{ReLu} $\rightarrow$ \texttt{1 Conv2D (5x5)} layers in Fig. ~\ref{fig:architecture} omitted.
The results for this condition are denoted as \texttt{NMP - P} in Table~\ref{tab:results}.
We first see that the constraint introduced by $Y_p$ consistently improves \texttt{Acc} across all datasets, however the effect on \texttt{Fno} and \texttt{F} is mixed; there is no significant difference for GuitarSet, Slakh and Phenicx, $Y_p$ improves performance slightly for MAESTRO, and degrades slightly for Molina. 
This suggests that even if the additional supervision is neutral for onset/offset detection, it is helpful for identifying note pitches, and we get the benefit of an additional output which contains some information about ornamentation and expressivity.

\subsection{Comparison with instrument-specific approaches}

\begin{table}[ht]
    \centering
    \includegraphics[width=0.9\columnwidth]{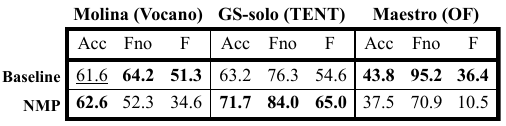}
    % \vspace*{-7mm}
    \caption{\small Average note event metrics on NMP vs. instrument-specific models for vocals, guitar and piano. The best score for each column is in bold.
    The compared instrument-specific model name is indicated in the column headers in parenthesis.
    All non-underlined results are statistically significantly different with $p<0.05$ compared with NMP via a paired t-test.}
    \label{tab:results-inst}
\end{table}

We've seen that the proposed model outperforms a comparable instrument-agnostic baseline on a variety of datasets.
To further understand the upper limits of our model, we provide a comparison with recent, open-source instrument-specific models.
\textbf{Onsets and frames (OF)}\cite{hawthorne_enabling_2019} is a polyphonic piano transcription method trained on the MAESTRO dataset which jointly predicts onset and note posteriorgrams using a CNN and RNN consisting of approximately 18M parameters, followed by a note-creation post-processing phase.
\textbf{Vocano} \cite{hsu_vocano} is a monophonic vocal transcription method which first performs vocal source separation, then applies a pre-trained pitch extractor followed by a note segmentation neural network, trained on solo vocal data. \textbf{TENT} \cite{su_tent_2019} is a monophonic solo guitar transcription method, which first performs melody contour extraction followed by playing technique detection of common guitar elements such as string bend, slide and vibrato using a CNN architecture, and a post-processing phase which obtains the final notes given the melody contour and the identification of the different playing techniques at each time frame.
We therefore only report results on the solo, monophonic half of GuitarSet.

For guitar, NMP outperforms TENT for all metrics, and more importantly,  these are state of the art results on GuitarSet to the best of our knowledge.
For vocals (\textit{Molina}), Vocano outperforms NMP in \texttt{Fno} and \texttt{F}, but the frame-level pitch accuracy (\texttt{Acc}) is comparable to NMP, suggesting that \texttt{Fno} could increase with improved onset detection. % in monophonic vocals. 
The largest difference in performance between NMP and an instrument-specific method is in the MAESTRO dataset compared with OF, which was specifically trained for piano transcription, and achieves 95.2\% \texttt{Fno}, in comparison to 70.9\% of our method (which is notably still a reasonably high score for this task).
The main reason for the difference in performance seems to be due to the onset detection accuracy which is higher in OF, since \texttt{Acc} is more similar for both methods (42.8\%  for OF vs. 37.5\% for NMP).
It is interesting to note that NMP would perform competitively in comparison with Melodyne\footnote{version 4.1.1.011, \url{http://www.celemony.com/en/melodyne}} on piano data according to the results obtained in  \cite{hawthorne_onsets_2018}, even if a direct comparison would not be possible since they reported results on another similar piano dataset.

\subsection{MPE Baseline}
Here we briefly validate how NMP performs at MPE, comparing NMP's MPE outputs with the output of the deep salience model\cite{bittner_deep_2017}. 
We report results on the Bach 10~\cite{duan2010multiple} and Su~\cite{su2015escaping} datasets, each of which contain 10 recordings of polyphonic western classical chamber music ensembles.
The MPE outputs for NMP outperform deep salience for the Bach10 dataset with a frame-level accuracy of $72.5 \pm 3.8$ versus $55.7 \pm 2.9$ for deep salience. 
However, deep salience achieves better results on Su $43.6 \pm 7.9$ where NMP gets $37.7 \pm 15.4$. 
While this is a small-scale validation, these results indicate that the information captured by $Y_p$ is meaningful and potentially competitive with strong baseline models. 
While the 3-bin-per-semitone resolution posteriorgrams may seem relatively low-resolution for this task, they can be used to estimate continuous multi pitch estimates, by using the amplitude values of the estimated $f_0$ bin, and those of its neighboring bins in frequency.
Note that despite not being trained on multi-instrument mixtures, it seems to achieve compelling results.

\subsection{Efficiency}

To illustrate the computational efficiency of NMP, we compare peak memory usage and total run time against MI-AMT.
Benchmarks were conducted on a 2017 Macbook Pro with a 3.1GHz Quad Core Intel Core i7 CPU and 16GB 2133MHz LPDDR3 Memory. 
All benchmarks were measured using first a ``short'' (.35 second) file of white noise to approximate the system's overhead, and a ``long'' (7 minute 45 second) file from the Slakh dataset in order to show a more realistic input for each method. Audio files were resampled to the expected sampling rate of the method before measurement.
We find that both methods are comparable in estimated overhead, with NMP using 490 MB peak memory and taking 7 s and MI-AMT using 561 MB and taking 10 s; however on the long file, NMP substantially outperforms MI-AMT, using only 951 MB peak memory and taking 24 s, while MI-AMT used 3.3 GB and took 96 s.
It's interesting to note that the peak memory of the instrument-specific models is even higher, with OF using 5.4 GB and Vocano using 8.5 GB.

\section{Conclusions}
We demonstrate that the proposed low-resource neural network-based model (NMP) can be successfully applied to instrument-agnostic polyphonic note transcription and MPE.
NMP outperforms a recent strong baseline note estimation model across five different datasets, and performs similarly to deep salience for MPE.
Further, we see that the use of harmonic stacking allows our model to remain low-resource while maintaining its performance. When compared with instrument-specific models, we see that NMP achieves state-of-the-art results on GuitarSet.
It however did not outperform the instrument-specific models for piano and vocals. 
Nevertheless, NMP has the benefit of being a ``one-size-fits-all'' solution, and has much lower computational requirements.
We hope to encourage further research into low-resource, multi-purpose AMT systems and believe that the proposed solution can be a valuable baseline. 

Future work could explore low-resource transcription of audio mixtures containing many instruments, and the use of offset predictions in this low-resource setting.
The note event creation method proposed is based on heuristics, and more carefully designed models similar to ~\cite{benetos2017polyphonic,ycart2018polyphonic} would likely result in note-event creation improvements.
While this work aimed to create a lightweight model from the start, we did not explore classic model pruning or compression techniques, which would further improve the efficiency. 
Finally, the interaction between the note and multipitch outputs could be explored, for example, to estimate note-level pitch bends.

\bibliographystyle{IEEEbib}
\bibliography{references}

\end{document}